\documentclass[10pt]{article}
\usepackage{amssymb,amsfonts,amsmath,latexsym}
\newcommand{\be}{\begin{equation}}
\newcommand{\ee}{\end{equation}}
\newcommand{\bea}{\begin{eqnarray}}
\newcommand{\eea}{\end{eqnarray}}
\newcommand{\ba}{\begin{array}}
\newcommand{\ea}{\end{array}}
\newcommand{\bt}{\begin{tabular}}
\newcommand{\et}{\end{tabular}}

\newcommand{\sump}{\mathop{{\sum}'}}

\newcommand{\fr}{\frac}
\newcommand{\ci}{\cite}
\newcommand{\cl}{\centerline}
\newcommand{\bs}{\bigskip}

\newcommand{\vs}{\vspace}

\newcommand{\en}{\eqno}

\newcommand{\bbib}{\begin{thebibliography}}
\newcommand{\ebib}{\end{thebibliography}}

\newcommand{\und}{\underline}
\newcommand{\lrar}{\leftrightarrow}

\begin{document}
\bs
\cl{\bf EXACT RESULTS FOR CONDUCTIVITY}
\cl{\bf OF 2D ISOTROPIC HETEROPHASE SYSTEMS}

\bs

\cl{\bf S.A.Bulgadaev \footnote{e-mail: sabul@dio.ru}}

\bs
\cl{Landau Institute for Theoretical Physics}
\cl{Chernogolovka, Moscow Region, Russia, 142432}

\bs

\begin{quote}
\footnotesize{
The duality relation for the effective conductivity $\sigma_{e}$
of 2D isotropic  heterophase
systems is used for obtaining the exact results for $\sigma_{e}$
at arbitrary number of phases $N.$
The exact values of $\sigma_{e}$ correspond to the fixed points
of the duality transformations.
The new exact results for $\sigma_{e}$, generalizing the well-known exact
values of $\sigma_{e}$ for  $N = 2,3,$ are found. It is shown that for
$N \ge 3$ there exist the whole hyperplanes
in the space of phase concentrations, on which $\sigma_{e}$ takes constant
values. These results are checked in the framework of
various approximations for different random heterophase systems.
}
\end{quote}

\bs
\cl{PACS: 72.80.Ng, 72.80.Tm, 73.61.-r}
\bs

\underline{1. Introduction}

\bs
The properties of the electrical transport of
classical inhomogeneous (randomly or regularly) heterophase systems,
consisting of $N \;(N \ge 2)$  phases with different
conductivities  $\sigma_i \; (i =1,2,...,N),$ have a great significance
for practice.
The main problem in theory of the electrical transport in these
systems is a calculation of the effective conductivity $\sigma_{e}$ at
arbitrary conductivities and phase concentrations \ci{1}.
Despite of seeming simplicity of this problem only a few exact results
have been obtained so far and all of them for two-dimensional systems \ci{2,3}.
They are based on the exact duality relation, which is a consequence of the
well-known duality between potential and tangent (or divergent free) fields
in two dimensions. This very important relation has been  obtained
firstly for isotropic two-phase systems, where it can be represented in the
form, connecting $\sigma_{e}$ at adjoint concentrations $x$ and $1-x$ \ci{2,3}
$$
\sigma_{e}(x,\sigma_1,\sigma_2) \sigma_{e}(1-x,\sigma_1,\sigma_2) =
\sigma_1 \sigma_2.
\en(1)
$$
It allows to find the exact universal,
remarkably simple, formula for $\sigma_{e}$ at equal phase concentrations
$x_1 = x_2 = 1/2$
$$
\sigma_{e} = \sqrt{\sigma_1 \sigma_2}.
\en(2)
$$
Analogous exact result for $\sigma_e,$ coinciding with (2), has been
obtained in $N=3$ case by Dykhne \ci{3}, when
$\sigma_1 \sigma_2 = \sigma_3^2$
and the concentrations of two first phases are equal $x_1 = x_2 = x.$
These exact values of $\sigma_e$ turned out to be very useful for various
applications of the theory, especially the equation (2), which has been
implemented for checking of various approximate expressions for $\sigma_e$
and in the percolation theory \ci{5,6}.
The dual relation (1) admits the generalizations on anisotropic systems
and polycrystals, systems in magnetic field  \ci{2,3,4}, but it cannot give
exact results for them (except isotropic systems in magnetic field \ci{4}).

In this letter, using a general duality relation, we will obtain exact results
for $\sigma_e$ of 2D isotropic
heterophase systems with arbitrary number $N$ of phases. They generalize
the above results for $N=2,3.$ In particular, it will be shown that for
$N>2,$ contrary to the $N=2$ case, there exist the whole hyperplanes in
the space of concentrations, on which $\sigma_e$ takes constant values.
The obtained results are checked in the framework of various approximate
schemes.

\bs
\underline{2. Symmetry, duality and  fixed points}

\bs

We begin with a discussion of some general properties
of the effective conductivity $\sigma_{e}$ of $N$-phase
systems with the partial conductivities $\sigma_i \ge 0\; (i = 1,2,...,N)$
and concentrations $x_i,$ which satisfy the normalization condition
$\sum_{i=1}^N x_i =1.$
The effective conductivity must be
a symmetric function of pairs of arguments ($x_i,\sigma_i$) and
a homogeneous (a degree 1) function of $\sigma_i.$
For this reason it is invariant under all permutations
$$
\sigma_{e}(x_1,\sigma_1|x_2,\sigma_2|...|x_N, \sigma_N) =
\sigma_{e}(P_{ij}(x_1,\sigma_1|x_2,\sigma_2|...|x_N, \sigma_N)),
$$
where $P_{ij}$ is the permutation of the i-th and j-th pairs of arguments,
and can be represented in the next form
$$
\sigma_{e}(x_1,\sigma_1|x_2,\sigma_2|...|x_N, \sigma_N) =
\sigma_s f(x_1,z_1|x_2,z_2|...|x_n,z_N).
\en(3)
$$
Here $z_i = \sigma_i/\sigma_s,$ where $\sigma_s$ is some normalizing
conductivity. For our purposes it will be convenient to choose
it in the symmetrical form $\sigma_s = (s_N)^{1/N},$ where $s_N = \prod_{i=1}^N \sigma_i$
is the N-th elementary symmetrical function. Then
$$
z_i =\sigma_i/(s_N)^{1/N},\quad \prod\nolimits_{i=1}^N z_i =1.
$$
As a result
the function $f$ is a symmetrical function of $N$ pairs $(x_i, z_i),$
only $N-1$ of them are independent.

The effective conductivity of $N$-phase systems must satisfy the following
natural boundary conditions: 1) $\sigma_e$ of $N$-phase system with
$n \; (2 \le n \le N)$ equal partial conductivities must reduce to
$\sigma_e$ of system with $N-n+1$ phases and
the concentrations of the phases with equal conductivities must add;
2) it has not depend on partial $\sigma_i$ and must reduce to the effective
conductivity of $(N-1)$-phase system, if the concentration of this phase
$x_i = 0$;
3) it must reduce to some partial $\sigma_i$, if $x_i = 1.$

A duality relation for $\sigma_{e}$ exists for any two-dimensional $N$-phase
system (see, for example, \ci{7,8}).
Its most general form for $\sigma_{e}$ valid at arbitrary phase concentrations
is
$$
\sigma_{e}(x_1,\sigma_1|...|x_N,\sigma_N)
\sigma_{e}(x_1,{\sigma_0}^2/\sigma_1|...|x_N,{\sigma_0}^2/\sigma_N) =
{\sigma_0}^2.
\en(4)
$$
where $\sigma_0$ is some arbitrary constant conductivity, characterizing
a duality transformation.

It is useful to note that the relation (4) can  be written also in the
other form
$$
\sigma_{e}(x_1,{\sigma_0}^2/\sigma_1|...|x_N,{\sigma_0}^2/\sigma_N) =
{\sigma_0}^2/
\sigma_{e}(x_1,\sigma_1|...|x_N,\sigma_N),
\en(4')
$$
which means that $\sigma_{e}$ at the inverse values of the partial
conductivities $\sigma_i$ is equal to the inverse $\sigma_{e}$.
For this reason it can be named also the inversion relation for
effective conductivity.

The mutually inverse values are related: their product is
equal $\sigma_0^2$ (Fig.1a). We will use also, for brevity,
the notations of the form $\{x\}$ for all whole sets of variables
like $x_i \; (i=1,...,N).$

An existence of the exact duality (or inversion) relation is very important
property of two-dimensional inhomogeneous systems.
It can be used for the obtaining of exact values of effective conductivity.
For this one needs to know the properties of the partial inversion
transformations
$$
J^{(i)}_{\sigma_0}: \sigma_i \to \sigma_0^2/\sigma_i, \quad
(J^{(i)}_{\sigma_0})^2 =I \quad (i=1,2...,N),
$$
where $I$ is the identity operator.
Each partial inversion transformation acts only on the partial
conductivity of the corresponding phase. Strictly speaking, the inversion
transformation is well defined
only for all $\sigma_i \ne 0,$ but one can extrapolate it formally on the case
$\sigma_i = 0.$ In this case it transforms a system with $\sigma_i = 0$ (a
dielectric phase) into a system with  $\sigma_i = \infty$ (an
ideal conducting phase).

The transformation  $J^{(i)}_{\sigma_0}$  has only one fixed point
$\sigma_i = \sigma_0.$ The i-th and j-th partial inversions,
having the same parameter
$\sigma_0^2 = \sigma_i \sigma_j$ (we denote them $J^{(i)}_{ij}$ and
$J^{(j)}_{ij}$ respectively), are very important, because they change
$\sigma_i$ into $\sigma_j$ and vice versa
$$
J^{(i)}_{ij} \sigma_i  = \sigma_i \sigma_j/\sigma_i = \sigma_j,\quad
J^{(j)}_{ij} \sigma_j = \sigma_i \sigma_j/\sigma_j =
\sigma_i.
\en(5)
$$
The physical sense has only the product of all partial inversions
with the same parameter $\sigma_0$
$$
J_{\sigma_0} = \prod_{i=1}^N \; J_{\sigma_0}^{(i)}, \quad  J_{\sigma_0}^2 = 1.
\en(6)
$$
Giving to $\sigma_0$ different
values, one can obtain various relations, connecting $\sigma_{e}$ at
different values of its arguments, which can serve
for checking of solutions found by other methods.

Since in our problem $J_{\sigma_0}$ acts on a space of
symmetrical homogeneous functions it is useful to consider its action on the
basic elementary symmetrical functions
$s_k(\{\sigma\}) \quad (\{\sigma\} = (\sigma_1,...,\sigma_N))$
$$
s_k = \sum^N_{i_1<i_2<...<i_k} \prod_{i_1}^{i_k} \sigma_{i_l} \quad
(k=1,...,N).
\en(7)
$$
It is easy to see that
$$
J_{\sigma_0} s_k = \sigma_0^{2k} s_{N-k}/s_N.
\en(8)
$$
It follows from (8) that only $s_N$ transforms into itself.
Since $\sigma_e$ is a homogeneous functions of degree 1,
it will be convenient to pass (for $\sigma_i \ne 0$) to the
projective variables $z_i$ and to
introduce projective basic symmetrical functions $\hat s_k$
$$
\hat s_k = s_k(\{z\}) = s_k/(s_N)^{k/N}, \quad k=1,.., N-1, \quad \hat s_N =1.
\en(9)
$$
Then we have from (9)
$$
J_{\sigma_0} z_i = 1/ z_i, \quad
J_{\sigma_0} \hat s_k = \hat s_{N-k},
\en(10)
$$
i.e. the inversion $J_{\sigma_0}$ for any $\sigma_0$
inverses $z_i$ and
interchanges adjoint $\hat s_k.$ This property strongly simplify a
search of possible fixed points needed for obtaining the exact equations for
$\sigma_e.$ In particular, the duality relation for functions $f$ has the
next form
$$
f(\{x\},\{z\})f(\{x\},\{1/z\}) = 1,
\en(4')
$$

We use the general duality relations (4,4') for the obtaining an exact
closed equation for $\sigma_e$ at some special values of $\sigma_{i}$
and $x_i.$  It follows from (4) that such equation can be obtained if one
finds a transformation $T(J,P)$, consisting  from inversions and permutations,
 and a combination of arguments $(\{x'\}, \{\sigma'\})$, which is a
fixed point of $T$
$$
T(J,P) \sigma_{e}(\{x'\},\{\sigma'\}) =
\sigma_{e}(\{x'\},\{\sigma'\}).
\en(11)
$$
Then, implementing this transformation to $\sigma_e(\{x'\},\{\sigma'\})$
in (4), one obtains the following exact equation
for $\sigma_e$ at these arguments
$$
\sigma_{e}^2(\{x'\},\{\sigma'\}) = \sigma_0^2,
\en(12)
$$
which gives an exact value of $\sigma_e$
$$
\sigma_e = \sigma_0.
\en(13)
$$
Here $\sigma_0$ is a parameter of the inversion, entering in $T(J,P).$
The corresponding exact equation for function $f$ has the form
$$
 f^2(\{x'\},\{z'\})=1,
\en(14)
$$
i.e. the exact value of $f$ at any fixed point is equal
to 1. Consequently, the possible exact values for $\sigma_e$ at these fixed
points must satisfy the next equalities
$$
\sigma_e(\{x'\},\{\sigma'\}) = \sigma_0 = (s_N)^{1/N}.
\en(15)
$$
If $T(J,P)$ contains the interchanging inversion $J_{ij}$ then
$$
\sigma_e = \sigma_0 = \sqrt{\sigma_i\sigma_j} = (s_N)^{1/N}, \quad i \ne j.
\en(15')
$$
Thus, (15,15') give a general form of the exact values of $\sigma_e$ in
these fixed points. The last equalities in (15) and (15') give a strong
constraint on possible fixed point values of $\sigma_i.$

In case of $N=2$ such fixed point exists at arbitrary $\sigma_i \; (i=1,2),$
but only at the equal phase concentrations $x_1=x_2=1/2$,
when the interchanging inversion $J_{12}$
is compensated by the permutation of the pairs of arguments.
It gives the known Keller -- Dykhne result (2).
The similar fixed point in case of $N=3$ was found by Dykhne \ci{3}
at equal phase concentrations ($x_i=x \; (i=1,2)$) and the additional
constraint on $\sigma_3$
$$
\sigma_3^2 = \sigma_0^2 = \sigma_1 \sigma_2,
\en(16)
$$
which remains $\sigma_3$ unchanged under the duality transformation.
The corresponding exact value for $\sigma_e$ coincides with (2) in accordance
with (15').

\begin{figure}
\begin{picture}(250,120)
\put(50,30){\vector(1,0){90}}
\put(50,30){\vector(0,1){90}}
\qbezier(130,40)(60,40)(60,110)
\put(145,30){$\sigma_1$}
\put(55,120){$\sigma_2$}
\put(75,5){(a)}
\put(120,0){\begin{picture}(100,50)%
\put(170,50){\vector(1,0){20}}
\put(120,100){\vector(0,1){20}}
\put(100,30){\vector(-1,-1){20}}
\put(100,30){\line(1,1){45}}
\put(170,50){\line(-3,1){60}}
\put(120,100){\line(1,-3){20}}
\put(170,50){\line(-1,1){50}}
\put(170,50){\line(-3,-1){75}}
\put(120,100){\line(-1,-3){25}}
\put(195,50){$x_2$}
\put(125,120){$x_3$}
\put(70,15){$x_1$}
\put(170,40){$1$}
\put(110,100){$1$}
\put(85,25){$1$}
\put(130,5){(b)}
\end{picture}}
\end{picture}

\vs{0.5cm}

{\small Fig.1. (a) A schematic plot of possible values of
$\sigma_i, \; (i=1,2),$ satisfying the relation
$\sigma_1\sigma_2 = \sigma_0^2$ with some fixed
value $\sigma_0$;
(b)  a concentration phase diagram for $N=3$, the bisectrisses correspond
to the pairwise equal phase concentrations.}
\end{figure}

It means that at these values of partial conductivities $\sigma_e$ is fixed
and does not depend on $x!$ Note that this value of $\sigma_e$
ensures the right values in the limits $x\to 1/2$ and $x\to 0,$ when they
must be $\sqrt{\sigma_1\sigma_2}$ and $\sigma_3$ respectively.

Of course, one can choose the pairs with equal concentrations by various ways.
For example, if  the phases $1$ and $3$ have the equal concentrations
$x$ and $\sigma_0^2 = \sigma_1 \sigma_3 = \sigma_2^2,$ then one obtains
$$
\sigma_{e} = \sqrt{\sigma_1 \sigma_3} = \sigma_2.
\en(17)
$$
Analogously, when $x_2 = x_3$ and
$\sigma_0^2 = \sigma_2 \sigma_3 = \sigma_1^2,$ one obtains
$$
\sigma_{e} = \sqrt{\sigma_2 \sigma_3} = \sigma_1.
\en(18)
$$
The corresponding lines of equal concentrations of different phases
on the concentration phase diagram are shown on Fig.1b,
where they coincide with the bisectrisses of the triangle (this triangle
is a concentration space of $N=3$ systems due to the normalization condition
$\sum_1^3 x_i=1$).
Their unique intersection point corresponds to the case of equal
concentrations of all phases. It follows from above results that under the
conditions (16)-(18) $\sigma_e$ remains fixed and does not change along the
corresponding  lines!
It is worthwhile to note that now, contrary to the $N=2$ case, the exact
value of $\sigma_e$ at the point of equal phase concentrations
$x_1 = x_2 = x_3 =1/3$ cannot be define from the duality relation
in general case. It can be found only in the above fixed points and
its value depends on the choice of fixed point. All these values are different.
Due to the additional constraints on $\sigma_i$, the form of possible
exact values of $\sigma_e$ admits also a value of the conductivity of the
third phase. For this reason one can say that
a structure of the  fixed points and a form of corresponding exact values
are defined by the involved interchanging inversion transformations and
the additional constraints.

\bs

\und{3. General $N$-phase case}
\bs

We show now that the duality relation (4) for $N$-phase systems
admits a generalization of these results on the $N$-phase systems.

Of course, for systems with $N > 3,$ there are similar  fixed points,
when $x_i = x_j$ and all other $\sigma_l \; (l\ne i,j)$ are equal between
themselves. But this case is trivial, since it reduces to the 3-phase case.
Fortunately, the new possibilities appear for $N>3.$

Let us consider firstly a case when $N$ is even $N=2M.$
Then a new fixed point is possible, which
corresponds to $M$ pairs of phases with equal concentrations
and equal products of the corresponding conductivities. For example,
the fixed point
$$
x_{2i-1}=x_{2i}=y_i, \quad
\sigma_0^2 = \sigma_{2i-1} \sigma_{2i} \quad (i=1,...,M)
\en(19)
$$
is possible. But all $\sigma_i$ in the phase pairs must be different.
In other words the conductivities of these pairs correspond to $M$
different  points on the curve (see Fig.1a)
$$
\sigma_0^2 = \sigma_{1} \sigma_{2}.
$$
The effective conductivity has in this case the next exact value
$$
\sigma_{e} = \sigma_0 = \sqrt{\sigma_{2i-1} \sigma_{2i}} \quad
(i=1,...,M).
\en(20)
$$
Note that the equal concentrations of the phase pairs $y_a \; (a=1,...,M)$
can be arbitrary, except the normalizing condition $\sum_{a=1}^M 2y_a =1.$
The exact value (20) does not depend on $y_a$ and ensures again the correct
values for $\sigma_e$ in the limits $y_a \to 0, \; 1 \; (a=1,...,M).$
Of course, the similar fixed points exist for other ways of the partition
of $N$ phases into the pairs  with equal concentrations.
A number of such points is equal $\#_{2M} = (2M-1)!!.$
The exact values of $\sigma_e$ in these points have always the same general
form
$$
\sigma_{e} = \sigma_0 = \sqrt{\sigma_{i} \sigma_{j}} \quad
(i<j, \; i,j = 1,...,N),
\en(21)
$$
where possible pairs $ij$ correspond to the phases with equal concentrations
in the corresponding fixed points.

Now we consider a case when $N$ is odd $N=2M+1.$ In this case a new fixed
point is possible if the above conditions of the even case (19) are
supplemented by a condition on $\sigma_{2M+1}$ analogous to the Dykhne
condition for $N=3$ case
$$
\sigma_{2M+1}^2	=	\sigma_0^2 = \sigma_{2i-1} \sigma_{2i} \quad (i=1,...,M).
\en(22)
$$
The corresponding effective conductivity  has again the exact value (20),
coinciding with the exact value of system with $N=2M$.
Of course, the other fixed points related with various ways of choice of
the phase pairs with equal concentrations are possible.
Their number is equal to $\#_{(2M+1)} = (2M+1)\#_{2M} = (2M+1)!!.$
A general form of the exact values of $\sigma_e$ remains the same as in the
even case with an additional equality
$$
\sigma_e = \sigma_0 = \sigma_{k} = \sqrt{\sigma_{i} \sigma_{j}}, \quad
i\ne j \ne k, \;\; i <j \quad (i,j,k = 1,...,N).
\en(23)
$$
where the pairs $ij$ correspond to the phases with equal concentrations
and $k$ corresponds to the unpaired phases.

It follows from (21,23) that the exact values of $\sigma_e$ at the point,
where all phase concentrations are equal, are not universal.

One can show that only the fixed points of type (19,22) and  their various
permutations are admissible for all dual transformations of the form
$T(J,P) = \prod J_a \prod P_b,$ where, for clarity, both products are assumed
to be finite, and the action of the product of inversions
$J=\prod J_a$ must be equal to (or compensated by) the product of permutations
$P=\prod P_b$ (we will not consider here possible, more sophisticated, forms
of $T$).
To do this it is useful to pass from partial conductivities $\sigma_i$ to
their logarithms $a_i = \ln \sigma_i, \; (-\infty \le a_i \le \infty).$
Then the partial inversion transformation $J^{(i)}_{\sigma_0}$ acts on $a_i$ as a
reflection $a_i \to A - a_i, \; A = 2\ln \sigma_0.$ The inversion
$J_{\sigma_0}$ acts on the set of variables $\{a\}$ in the following way
$$
a_i \to A - a_i, \; (i=1,...,N)
\en(24)
$$
The action of the product of $L$ different inversions
$J = \prod_{s=1}^L J_{\sigma_0^{(s)}}$ gives analogous result
$$
a_i \to A_L + (-1)^L a_i, \; (i=1,...,N)
\en(25)
$$
with $A_L = \sum_{s=1}^L (-1)^s A_s, \; A_s = 2\ln \sigma_0^{(s)}.$
Since all parameters of inversion transformations collect into one constant
$A_L,$ it will be enough to consider only $L=1$ or $L=2.$
In order to find any fixed point the result (25) must be equivalent to some
permutation $P$ of the set $\{a\}$
$$\{a\}_P = (a_{1_P},...,a_{N_P}).$$
Let us consider
firstly a case of even $L=2.$ Then one obtains from (25)
$$
a_{i_P} = A_2 + a_i, \; (i=1,...,N)
\en(26)
$$
The set of equations (26) is very restrictive. If any $a_i$ remains unchanged,
than $A_2 = 0$ and all other $a_i$ also remain unchanged. It means that
in this case $J=I.$ If (26) gives a permutation of any  pair
$a_i \lrar a_k, \; (i\ne k),$ than it follows from (26) that again $A_2 = 0.$
As is known any permutation $P$ decomposes into product of independent chains
of simple permutations. Then,
considering all possible chains similar to the abovementioned simplest case
of pair permutation, one can see that equations (26) take place only when
$A_2=0.$ In case of odd $L=1$ equations (25) take the form
$$
a_{i_P} = A_1 - a_i, \; (i=1,...,N)
\en(27)
$$
If (27) gives a permutation of some pair $a_i \lrar a_k, \; (i\ne k),$ then
one obtains for $a_i,a_k$ only one equality (since in this case both equations
coincide)
$$
a_i + a_k = A_1.
\en(28)
$$
In case of any larger chains of permutation one obtains additional
equalities, which have only trivial solution. For example, in case of
3-chain ($1\to 2 \to 3 \to 1$) they are
$$
a_1 + a_2  = a_2 + a_3 = a_1 + a_3 = A_1.
\en(29)
$$
It follows from (29) that this is possible only if
$a_1=a_2=a_3 =a, \; A_1 =2a,$ but this corresponds to the one phase case.
Consequently, the equations (27) can take place if and only if
the permutation $P$ decomposes into product of pair permutations (for $N=2M$)
or into product of pair permutations and one identical transformation
(for $N=2M+1$).
Then one obtains in case of pair permutations of type $a_{2i-1} \lrar a_{2i}$
the next equalities
$$
a_{2i-1} + a_{2i} = A_1 \quad (i=1,...,M), \quad (N=2M),
$$
$$
a_{2i-1} + a_{2i} = 2a_{2M+1} = A_1 \quad (i=1,...,M), \quad (N=2M+1).
\en(30)
$$
All other solutions can be obtain from (30) by various permutations.
These solutions give exactly the fixed points (19,22)
after returning to the variables $\{\sigma \}.$

\bs
\und{4. Checking of exact results for random heterophase systems}

\bs
The existence of the fixed point hyperplanes in space of
concentrations and a nonuniversality of the
corresponding exact values at the point of all equal concentrations
for all $N \ge 3$ can be confirmed for random heterophase systems in the
framework of corresponding approximate schemes \ci{5,9}.
One needs only that such approximate scheme satisfies the duality relation.

Firstly  we will show how these properties look like in the EM approximation
for $N$-phase square lattice random resistor network (RRN)
systems. This approximation gives the following equation for determination
of $\sigma_{e}$ \ci{5}
$$
\sum_1^N \; x_i \;\fr{\sigma_{e} - \sigma_i}{\sigma_{e} + \sigma_i} = 0.
\en(31)
$$
It can be written as an algebraic equation of the $N$-th order
$$
\sigma_{e}^N + a_1 \sigma_{e}^{N-1} + ... + a_{N-1} \sigma_{e} + a_N = 0,
\en(32)
$$
where
$$
a_k = s_k - 2 \bar s_k = -s_k +2\tilde s_k \quad  (k=1,...,N-1), \quad
a_N = - s_N.
\en(33)
$$
Here
$$
\bar s_k = \sum_{i=1}^N x_i \sigma_i
\sump_{i_1<i_2<...<i_{k-1}}^N \prod_{i_l=i_1}^{i_{k-1}} \sigma_{i_l}, \quad
\tilde s_k = \sum_{i=1}^N x_i \sump_{i_1<i_2<...<i_k}^N
\prod_{i_l=i_1}^{i_k}\sigma_{i_l},
$$
where  all
$\sump_{i_1<i_2<...<i_k}^N$ do not contain $i_l = i.$
The coefficients $a_k$ are the homogeneous functions of $\sigma_i$
of degree k and the linear functions of concentrations $x_i.$
For this reason they are related
only with the corresponding symmetrical functions $s_k$ and their "averages"
$\bar s_k$ or $\tilde s_k$ (since there are different ways to define averaged
$s_k$).
The eq. (32) must have one physical solution $\sigma_e \ge 0.$

It is easy to see
that $\sigma_{e}$ determined from the equation (32) satisfies the duality
(and inversion) relation (4), since
the equation (31) for the inverse conductivities has the next
property  (for all $\sigma_i \ne 0$)
$$
\sum_1^N x_i
\fr{\sigma_{e} -\sigma_0^2/ \sigma_i}{\sigma_{e} + \sigma_0^2/\sigma_i} = -
\sum_{i=1}^N x_i
\fr{\sigma_0^2/\sigma_{e} - \sigma_i}{\sigma_0^2/\sigma_{e} + \sigma_i}=0.
$$
Moreover, since $a_N = - s_N$, one can pass directly to
the similar equation for $f$ by dividing all its terms on $s_N$
$$
f^N + \hat a_1 f^{N-1} + ... + \hat a_{N-1} f - 1 = 0,
\en(34)
$$
where now the coefficients $\hat a_k$ are expressed through projective variables
$z_i$
$$
\hat a_k = a_k(\{z\}) = a_k/(s_N)^{k/N}, \quad k=1,...,N-1.
\en(35)
$$
Firstly we consider the case of equal concentrations $x_i= 1/N.$
Then one can see that the coefficients $\hat a_k$  and
$f$ must depend only on symmetrical functions $\hat s_k \; (k=1,...,N-1)$
$$
\hat a_k =(N-2k)/N \quad \hat s_k, \quad f=f(\hat s_1,..., \hat s_{N-1}),
\en(36)
$$
Due to the transformation rules of $\hat s_k$ (10) the transformed $f$ in the
duality relation (4) has the form
$$
J f(\hat s_1,..., \hat s_{N-1}) = f(\hat s_{N-1},...,\hat s_1).
\en(37)
$$
Thus, one must have at the fixed point the next equalities
$$
\hat s_{N-k} = \hat s_k, \quad k=1,...,N-1.
\en(38)
$$
It is easy to check that they are satisfied at the fixed points (19),(22) and
that the coefficients $\hat a_k$ satisfy at these points the equations
$$
\hat a_{N-k} = - \hat a_{k}.
\en(39)
$$
It follows from (39) that for even $N=2M$ the coefficient $\hat a_M =0.$
Consequently, one
has from (38),(39) that the polynomial in (34) can be represented in the
factorized form
$$
(f-1)P_{N-1}(f) = 0.
\en(40)
$$
where a polynomial $P_{N-1}(f)$ of degree $N-1$  has the next form
$$
P_{N-1}(f) = \sum_{i=1}^M f^{2M-i} \sum_{l=0}^{i-1}\hat a_l  +
f^{M-1} \sum_{l=0}^{M-1}\hat a_l  + (\sum_{l=0}^{M-2}f^l)
(\sum_{l=0}^{M-2}\hat a_l), \quad N=2M;
$$
$$
P_{N-1}(f) = \sum_{i=0}^M f^{2M-i} \sum_{l=0}^{i}\hat a_l +
(\sum_{l=0}^{M-1}f^l)(\sum_{l=0}^{M-1}\hat a_l), \quad N=2M+1.
\en(41)
$$
Since all coefficients of $P_{N-1}(f)$ are positive, one can see from (40)
and (41) that equations (32) and (34) have always only
one physical solution at any fixed point (19),(22)
$$
f=1, \quad \sigma_e = (s_N)^{1/N} = \sqrt{\sigma_i \sigma_j} \quad (i\ne j).
\en(42)
$$
which coincides with the corresponding exact values from (21),(23).

Now we pass to the case of pairwise equal concentrations. Unfortunately,
a complete analysis in this case is rather complicated and, at the moment,
it is not done in general form.
For this reason  we have considered only the systems with $N=4,5,6.$ It is
convenient to consider the odd and even cases separately.
We begin with the even case $N=4,6.$ Due to apparent
permutation symmetry $S_N$, it will be enough to
consider only fixed points (19).
For $N=4,6$ at the fixed points
$$
x_1 = x_2 =y_1, \quad x_3 = x_4 =y_2, \quad  \sum_{a=1}^2 y_a =1/2, \quad
\sigma_1\sigma_2 = \sigma_3\sigma_4 = \sigma_0^2,
\en(N=4)
$$
$$
x_{2i-1} = x_{2i} = y_i,  \quad  \sum_1^3 y_i = 1/2, \quad
\sigma_{2i-1}\sigma_{2i} = \sigma_0^2, \quad (i=1,2,3),
\en(N=6)
$$
one can show by direct calculation, using equalities (38), that for all
$y_i$
$$
\hat a_k = -\hat a_{N-k}, \quad k=1,2, \quad  \hat a_2=0
\en(N=4),
$$
$$
\hat a_k = -\hat a_{N-k}, \quad k=1,2,3, \quad  \hat a_3=0
\en(N=6).
$$
Thus the polynomial in equation (34) reduces also to the
factorized form
$$
(f-1)P_{N-1}(f,y) = 0,
\en(43)
$$
where the polynomials $P_{3,5}(f,y)$ have the form  (41), but now the
coefficients $\hat a_k$ depend on concentrations $y.$
Due to factorization, eqs.(32),(34) have the required solutions of the form
(20).

In the odd case $N=2M+1$ we have checked for $M=1,2$ that the necessary
conditions $\hat a_k = -\hat a_{N-k}, $
are satisfied at the fixed point (22).
As a result the polynomial in the defining eq.(34) reduces also to the
factorized form
$$
(f-1)P_{2M}(f,y) = 0,
\en(44)
$$
which means that in the odd cases the EM approximation has also the fixed
point hyperplanes in the concentration space and that at these fixed points
there are always the solution,
which reduces at these points to the corresponding exact values (23).

We are sure that the EM approximation for general $N$ contains all fixed
points (19,22) with the corresponding exact values.

Now we check the obtained exact results in framework of two other
approximations applicable for some random heterophase systems.
They were introduced firstly for two-phase systems in \ci{9} and
generalized later on $N$-phase
systems in \ci{10}. Fortunately, they give for $\sigma_e$ the expressions
more simple than the EM approximation.

We begin with the finite maximal scale average (FMSA) approximation valid
for random systems with compact inclusions. The corresponding approximate
expression for $\sigma_e$ of $N$-phase systems is \ci{9,10}
$$
\sigma_e(\{x\},\{\sigma\}) = \prod_{i=1}^N \sigma_i^{x_i}.
\en(45)
$$
It is easy to check that (45) reproduces at the corresponding fixed points
the  exact values from (20),(21),(23) due to the normalization condition for
concentrations
$$
\sigma_e = \prod_{i=1}^M \sigma_0^{2y_i} = \sigma_0 = (s_N)^{1/N},
\en(N=2M)
$$
$$
\sigma_e = \sigma_0^{1-2\sum_1^M y_i} \prod_{i=1}^M \sigma_0^{2y_i} =
\sigma_0 = (s_N)^{1/N}.
\en(N=2M+1)
$$
It is interesting that in this case $\sigma_e$ at the point of all equal
concentrations $x_i=1/N$ always coincide with $s_N^{1/N},$ (i.e. for arbitrary
$\sigma_i.$

Another approximate expression for $\sigma_e$ is valid for the composite
random heterophase models \ci{9}. It has the following form in $N$-phase
case \ci{10}
$$
\sigma_{e}(\{x\},\{\sigma\}) = \sqrt{\langle \sigma \rangle
\langle \sigma^{-1} \rangle^{-1}} =
\left(\fr{s_N \bar s_1}
{\tilde s_{N-1}}\right)^{1/2}.
\en(46)
$$
By direct substitution of the fixed points values (19),(22) into (46)
one can also reproduce the exact values from (20),(21),(23).

\bs
\und{5. Conclusion}
\bs

Using the exact duality relation, symmetry and inversion properties of
2D isotropic heterophase systems we have found all possible exact values
of their effective conductivities.
The obtained results take place for various heterophase systems (regular and
nonregular as well as random), satisfying the duality relation (4), and show
very interesting properties. Especially, an existence of hyperplanes in the
concentration space with  constant $\sigma_e$ is unusual.
It will be very desirable to check them experimentally.

\bs
\und{ Acknowledgments}
\bs

This work was supported by the RFBR grants \# 2044.2003.2 and \# 02-02-16403.

\end{document}